\documentclass[twocolumn]{aastex631}

\usepackage{xspace}
\usepackage{xcolor}
\usepackage{graphicx}
\usepackage{soul}
\usepackage{enumerate}
\usepackage{mathrsfs}
\newcommand{\hxmt}{{HXMT}\xspace}

\newcommand{\nicer}{\textit{NICER}\xspace}

\newcommand{\fermi}{{\it Fermi}\xspace}
\newcommand{\swift}{{\it Swift}\xspace}

\newcommand{\nustar}{{\it NuSTAR}\xspace}

\newcommand{\oergs}[1]{$10^{#1}$ erg s$^{-1}$}
\newcommand{\rxj}{RX\,J0209.6$-$7427\xspace}
\newcommand{\rxjs}{RX\,J0209\xspace}

\tabletypesize{\footnotesize}

\defcitealias{WestEtal2017a}{{\tt WWB17}}

\begin{document}

\title{Theoretical Analysis of the RX\,J0209.6-7427 X-ray Spectrum During Giant Outburst}

\author[0009-0007-1219-7508]{Brent F. West}
\affiliation{Department of Physics, United States Naval Academy, Annapolis, MD 21402, USA}

\author[0000-0002-3718-1293]{Peter A. Becker}
\affiliation{Department of Physics and Astronomy, George Mason University, Fairfax, VA 20030, USA}

\author[0000-0003-3902-3915]{Georgios Vasilopoulos}
\affiliation{Department of Physics, National and Kapodistrian University of Athens, University Campus Zografos, GR 15784, Athens, Greece}
\affiliation{Institute of Accelerating Systems \& Applications, University Campus Zografos, Athens, Greece}

\begin{abstract}

We model the spectral formation occurring in the binary X-ray pulsar RX~J0209.6-7427 during the 2019 super-Eddington outburst. Using a theoretical model previously developed by the authors, we are able to produce spectra that closely resemble the phase-averaged X-ray spectra observed using \textit{NuSTAR} and \textit{Insight}-HXMT during low and high luminosity states of the outburst, respectively. The theoretical model simulates the accretion of fully ionized gas in a dipole magnetic field, and includes a complete description of the radiation hydrodynamics, matter distribution, and spectral formation. Type II X-ray outbursts provide an opportunity to study accretion over a large range of luminosities for the same neutron star. The analysis performed here represents the first time both the outburst low and high states of an accretion-powered X-ray pulsar are modeled using a physics-based model rather than standard phenomenological fitting with arbitrary mathematical functions. We find the outer polar cap radius remains constant and the column is more fully-filled with increasing luminosity, Comptonized bremsstrahlung dominates the formation of the phase-averaged X-ray spectrum, and a negative correlation exists between cyclotron centroid energy and luminosity, as expected. The super-Eddington nature of the outburst is rendered possible due to the low scattering cross section for photons propagating parallel to the magnetic field. We also find emission through the column top dominates in both the low and high states, implying the pulse profiles should have a roughly sinusoidal shape, which agrees with observed properties of ultra-luminous X-ray pulsars. 

\end{abstract}



\section{Introduction} 

Accretion-powered X-ray pulsars (XRPs) are intriguing systems that enable the study of complex radiative processes in the presence of strong magnetic fields \citep[for review see][]{Mushtukov2023}. 
The accreting material is entrained from the surrounding accretion disk by the strong magnetic field, and then channeled onto a magnetic pole of the neutron star (NS), forming an accretion column (AC). The luminosity of these systems range from $\sim$$10^{34}$ up to $\sim$\oergs{38}, and even above the Eddington limit in extreme cases. The classical spherical Eddington limit, $L_{\rm Edd}=1.25\times 10^{38}M/M_\odot\,{\rm erg\,s}^{-1}$, is the maximum luminosity a star of mass $M$ can have, in the Thomson scattering limit, and not eject the accreting material by radiation pressure, where $M_\odot$ is the mass of the sun.

Be X-ray binaries are especially interesting \citep[BeXRBs; for a review see][]{2011Ap&SS.332....1R}. The accretion disk from the companion Be star in these systems periodically provides enhanced mass transfer onto the NS. The variable mass accretion rate converts gravitational potential energy into a highly luminous release of radiation, designated as a Type~II, or giant, X-ray outburst \citep{2014ApJ...790L..34M}. The X-ray luminosity can exceed $\sim 10^{39}\,{\rm erg\,s}^{-1}$; hence, Type~II outbursts from BeXRBs provide a unique window into the behavior of the closest ultra-luminous X-ray sources \citep[ULXs, for a review see][]{KingEtal2023} and establish an important connection between typical XRPs and the ULX pulsars (ULXPs).

Application of theoretical models to XRPs in the super-Eddington regime represents an important step towards investigations of ULX systems. The high luminosities of some types of accretion-powered XRPs, and particularly the ULXPs, suggest radiation pressure plays a leading role in the accreting gas dynamics. Some sort of beaming may be required to preserve the structure of the surrounding accretion disk \citep{2001ApJ...552L.109K,KingEtal2023}. The luminosities of the ULXPs exceed $L_{\rm Edd}$, implying the accreting gas is unstable to ejection via the pressure of the intense radiation field \citep{KingEtal2023}. The classical Eddington limit assumes full Thomson scattering, however, which is not applicable in XRPs because the electron scattering cross section for photons propagating parallel to the magnetic field is several orders of magnitude smaller than the Thomson value \citep{Ventura1979,MeszarosAndVentura1979,BeckerEtal2012,2015MNRAS.454.2539M}.

Type~II outbursts offer a unique laboratory to perform a detailed analysis of the physical processes occurring in the AC over a large range of  $L_{\rm X}$. For low $L_{\rm X}$, the accretion is sub-critical, meaning that $L_{\rm X} \lesssim L_c$, where $L_c$ is the critical luminosity above which radiation pressure becomes dynamically important \citep[e.g.,][]{BeckerEtal2012}. The AC is believed to transition to the super-critical regime at the peaks of the most extreme Type-II outbursts, with luminosity $L_{\rm X} \gtrsim L_c$ \citep{2014ApJ...790L..34M}. This can significantly alter both the dynamical structure of the AC, as well as the dominant mode of spectral formation \citep{2013A&A...551A...1R}.

Physical simulation of the AC requires consideration of a variety of physical processes occurring in the magnetized, accreting plasma, as it passes through either a gas-dominated discontinuous shock (in low-luminosity sources) or a continuous (smooth) radiation-dominated shock (in high-luminosity sources such as the Type~II outburst studied here). Seed photons produced via bremsstrahlung, cyclotron, and blackbody emission are reprocessed via bulk and thermal Comptonization before escaping through the walls and top of the column. The first model to include all of these physical features was developed by \cite{BeckerAndWolff2007}, who applied their model to simulate the formation of the observed X-ray spectra for the sources Her X-1, Cen X-3, and LMC X-4.

The model of \cite{BeckerAndWolff2007}, and also the similar model of \cite{FarinelliEtal2016}, proved successful at reproducing the observed X-ray pulsar spectra in super-critical sources, where radiation pressure plays a dominant role in controlling the dynamics of the accreting material. Additional progress was made by \citet{BeckerWolff2022}, who developed a generalized conical geometry model with application to X-ray pulsars over a wide range of luminosities, and incorporates a more realistic bulk fluid velocity profile. However, these models do not treat the hydrodynamical structure of the column in detail; instead, they assume the velocity profile of the accreting gas is specified (known) in advance.

To date, the most self-consistent model for spectral formation in accretion-powered X-ray pulsars is the one developed by \citet{WestEtal2017a,WestEtal2017b}, which is based on an iterative method that yields simultaneous, coupled solutions for the radiation-hydrodynamical structure of the accretion column, along with the radiation distribution escaping through the walls and top of the column. We focus on the application of this model to simulate the Type~II outburst observed from \rxj (hereafter \rxjs) that occurred in 2019 \citep{2020MNRAS.494.5350V}.

The model of \citet[][hereafter {\tt WWB17}]{WestEtal2017a} is a radiation-hydrodynamics simulation cast in a dipole geometry in which the bulk flow velocity is computed by solving a one-dimensional hydrodynamical equation coupled with ion and electron energy equations that fully account for the energy exchange processes between ions, electrons, and photons. It was developed primarily for the study of luminous pulsars. In this paper, we focus on the application of the \protect\citetalias{WestEtal2017a} model to the analysis and interpretation of X-ray data obtained using \textit{NuSTAR} and \textit{Insight}-HXMT during the \rxjs outburst.

The strong magnetic fields in accretion-powered XRPs cause the cyclotron energy levels of the electrons to be quantized; therefore, the electron scattering cross section is a complex function of the photon energy and propagation direction \citep{Ventura1979,MeszarosAndVentura1979}.
The electron scattering cross section is also qualitatively different and depends on whether the photon is polarized in the ordinary or extraordinary mode. The extraordinary mode photons, in particular, have resonant interactions with the electrons which cause a sharp increase in the scattering cross section at the cyclotron centroid energy, $E_{\rm cyc}$, given by the standard relation
\begin{equation}
    E_{\rm cyc}(r) = 11.57\,n\,B_{12}(r) \ \rm{keV} \ ,
    \label{eqn:cycenergy}
\end{equation}
where $n$ corresponds to the Landau level from which the radiative de-excitation to the ground state occurs \citep{2019A&A...622A..61S}, and the dipole magnetic field strength varies with the radius, $r$, according to
\begin{equation}
    B_{12}(r)=\frac{B_*}{10^{12}\,{\rm G}}\left(\frac{r}{R_*}\right)^{-3} \ .
    \label{eqn:BfieldStrength}
\end{equation}
Here the stellar radius and the surface magnetic field strength at the polar cap are denoted by $R_*$ and $B_*$, respectively. 

The interaction of the extraordinary mode photons with the electrons leads to the formation of a cyclotron resonant scattering feature (CRSF) at photon energy $E_{\rm cyc}$, which is often observed as an absorption-like feature, or simultaneous multiple features, in the spectra of XRPs. They are not true absorption features, but rather represent the scattering of radiation out of the beam due to collisions with electrons in the accreting plasma. The implications of cyclotron scattering for the formation of the dual CRSFs during the giant outburst of \rxjs are discussed below.

\section{Overview of the {\tt WWB17} Model} \label{sec:model updates}

We investigated the radiation-hydrodynamics inside the AC of \rxjs using the iterative model of {\tt WWB17}, which is briefly described here. The complete process to build a self-consistent radiation and hydrodynamical pulsar AC model is described in detail in \citet{WestEtal2017a,WestEtal2017b}, in which the accretion column aligns with the magnetic field dipole shape, and all of the physical quantities are functions of the radial coordinate, $r$, measured from the center of the NS, meaning they are averaged across the column cross-section at a given value of $r$.

The range of the dipole magnetic field radial coordinates provides the computational domain for solving a set of five coupled ordinary differential equations, which describe the spatial evolution of the bulk flow velocity, the total energy flux, and the electron, ion, and radiation sound speeds; therefore, the ion and electron temperatures are determined using separate energy equations. Coulomb collisions in high-luminosity sources rapidly equilibrate the ion and electron temperatures, but they are less efficient in low-luminosity sources, and the two temperatures may be quite different. Simultaneous solutions of the five coupled differential equations yield a complete description of the dynamical and thermal structure inside the AC.

The bulk flow velocity and electron temperature solutions are subsequently used to solve a second-order elliptic partial differential photon transport equation to obtain the radiation distribution inside the AC, which includes terms describing advection, diffusion, free-free (bremsstrahlung) emission and absorption, cyclotron emission and absorption, bulk and thermal Comptonization, and the escape of radiation through the column walls and top. Blackbody seed photons from the surface of the thermal mound are also included in the model. 

An upper boundary condition is imposed at the column top where photon transport makes a transition from diffusion to free streaming at the last scattering surface, and a zero total energy flux boundary condition is imposed at the stellar surface. The scattering of the photons by the electrons in the accreting plasma is modeled using the cross sections $\sigma_{||}$ and $\sigma_\perp$ for photons diffusing parallel or perpendicular to the magnetic field lines, respectively. The model also implements an angle-averaged cross section, $\bar\sigma$, which is used to describe the Compton scattering interaction between the photons and electrons.

The solution for the radiation distribution inside the AC is used to compute the radial profile of the inverse-Compton temperature, $T_{\rm{IC}}$, which plays a critical role in controlling the thermal balance between the gas and the radiation field. The complete set of six conservation equations, comprising five hydrodynamical relations and one radiation transport equation, must be solved iteratively because the photon transport equation requires hydrodynamical information (electron temperature and bulk velocity profiles), whereas the hydrodynamical solutions require knowledge of the inverse-Compton temperature profile. The iterative process is considered converged when both the electron temperature and inverse-Compton temperature profiles each change by less than 1\% from the previous iteration. The result is a fully self-consistent model for the hydrodynamical and radiative structure of the AC.

\section{Astrophysical Application} \label{sec:application}

We used the \protect\citetalias{WestEtal2017a} model to analyze the formation of the phase-averaged X-ray spectra within 2 epochs during the giant outburst observed from \rxjs in 2019. Data were collected by \nustar and \textit{Insight}-HXMT during the low and high states, respectively, for both epochs as the NS was accreting at super-Eddington rates.

\begin{figure}

\resizebox{\hsize}{!}{
  
  \includegraphics[width=8.4cm,angle=0]{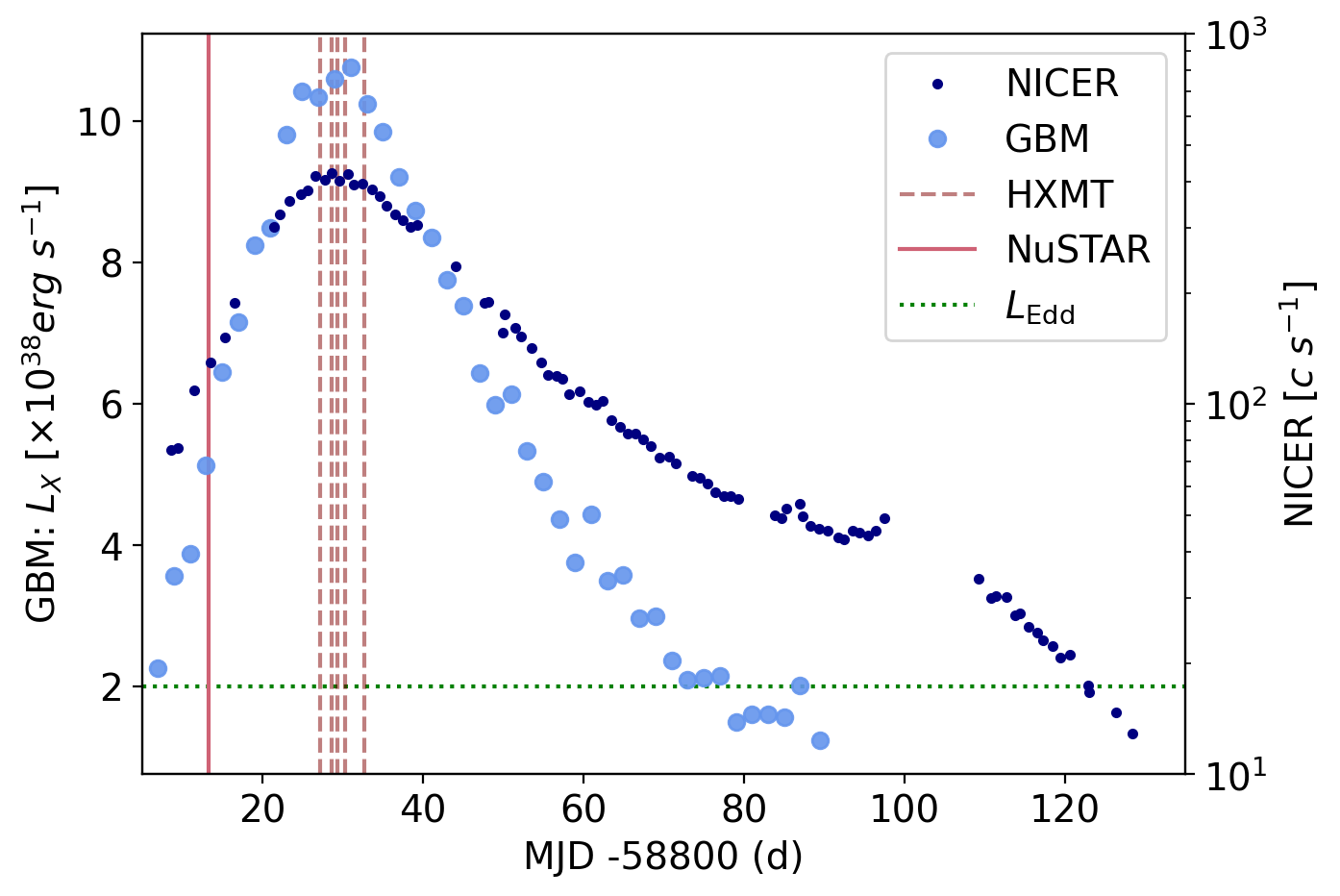}
  }

\includegraphics[width=7.7cm,angle=0]{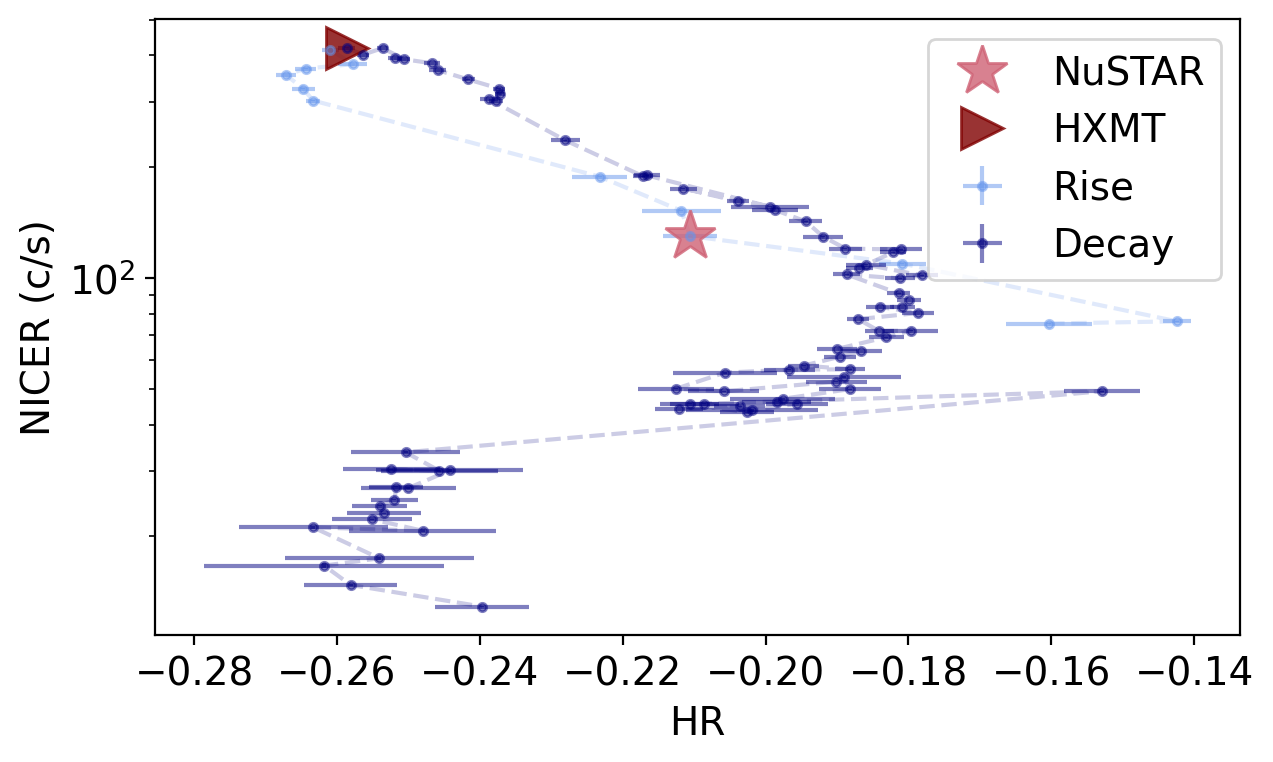}

  \caption{\emph{Top:} X-ray light curve of 2019 outburst of \rxjs, $L_{\rm X}$ is based on \fermi/GBM pulsed fluxes while secondary axes denotes \nicer count rates (0.3-8.0\,keV). Horizontal line marks the Eddington limit for a canonical NS based on Thomson scattering. \emph{Bottom:} \nicer Hardness-intensity diagram based on the R1: 1.0-2.0\,keV and R2: 2.0-4.5\,keV bands, where HR is defined as $(R2-R1)/(R2+R1)$.}
  \label{fig:LC}
\end{figure}

\subsection{2019 Outburst of RX J0209: Data Collection}

\rxjs is a BeXRB pulsar (9.3\,s period) located in the outer wing of the Small Magellanic Cloud (SMC). During its 2019 major outburst its observed $L_{\rm X}$ exceeded \oergs{39} \citep{2020MNRAS.494.5350V}. The outburst was monitored by all-sky X-ray instruments such as \fermi/GBM and \swift/BAT, and pointed observations were made with \nicer. 
An X-ray light curve of the outburst that covers the brightest phase is plotted in Figure~\ref{fig:LC}, where $L_{\rm X}$ is estimated from GBM pulsed fluxes (12\,\textendash\,50 keV) at a distance of $D = 55\,$kpc \citep[see][]{2023MNRAS.520..281K}.

The Nuclear Spectroscopic Telescope Array (\nustar) mission is the first focusing high-energy X-ray telescope operating in the 3\,\textendash\,79\,keV energy range \citep{2013ApJ...770..103H}. We analysed the \nustar DDT observation of \rxjs (obsid: 90502352002) following \citet{2020MNRAS.494.5350V}.
The Hard X-ray Modulation Telescope \citep[\hxmt;][]{2020SCPMA..6349502Z} collected data during 5 visits (proposal id: P0214059) over an $\sim$5 day period around the peak of the outburst \citep{2022ApJ...938..149H}.
\hxmt is equipped with the Low Energy (LE, 0.7\,\textendash\,15\,keV), Medium Energy (ME, 5\,\textendash\,30\,keV), and High Energy (HE, 20\,\textendash\,250\,keV) X-ray telescopes. The data were downloaded from the \hxmt archive and analysed with {\tt hxmtsoft v2.04} using default options.

Based on the collected statistics, we limited our study to the following energy ranges for each detector; 2.0\,\textendash\,8\,keV for LE, 8\,\textendash\,30\,keV for ME; and 30\,\textendash\,100\,keV for HE \citep[see also][]{2022ApJ...938..149H}. We validated our analysis by comparing with results in the literature \citep[i.e.][]{2022MNRAS.517.3354L,2022ApJ...938..149H}. Specifically, we fitted the same spectral models and we compared the model parameters and residuals obtained with the previous results. Finally, we used the unfolded spectra in the \citetalias{WestEtal2017a} modeling process. 

\subsection{Observed Luminosity and Mass Flow Rate}

The bolometric X-ray luminosity, $L_{\rm X}$, is related to the mass accretion rate, $\dot M$, via the standard formula
\begin{equation}
    L_{\rm X} = \frac{G M_* \dot M}{R_*} \ ,
    \label{eq:mdot}
\end{equation}
where we assume all gravitational potential energy is converted into radiation, which is a reasonable approximation given the extreme density of the NS crust. We do not have complete knowledge of the bolometric luminosity, but we can obtain a reasonable approximation by using the energy bands corresponding to the specific detectors used to make the X-ray observations. For the low state of the RX\,J0209 outburst observed in 2019 using {\it NICER} and {\it NuSTAR}, the bolometric X-ray luminosity, $L_{\rm X}$, in the energy range 0.5\,\textendash\,70.0\,keV, was found to be $L_{\rm X}=5.54 \times 10^{38}\,{\rm erg\,s}^{-1}$ \citep{2020MNRAS.494.5350V}. For the high state of the 2019 outburst, observed using {\it Insight}-HXMT, the bolometric luminosity in the energy range 1.0\,\textendash\,150.0\,keV was found to be $L_{\rm X} = 1.10 \times 10^{39}\,{\rm erg\,s}^{-1}$ \citep{2022ApJ...938..149H}. The luminosity values are computed assuming isotropic emission with a source distance $D = 55\,$kpc.

\subsection{Cyclotron Resonance Scattering Feature (CRSF)}

The formation of a CRSF in the observed X-ray spectrum via resonant cyclotron scattering is a challenging technical problem and beyond the scope of the present paper \citep[e.g.,][]{SchwarmEtal2017a,SchwarmEtal2017b}. The location of the imprint radius provides valuable information about the strength of the magnetic field, and remains an active research topic. The hydrodynamic equations of the {\tt WWB17} model are based upon a dipole magnetic field geometry, and converging the model with a predetermined imprint radius has fundamental effects on the column height and structure.

The supercritical luminosities for both the low and high states of the RX J0209 outburst suggest the CRSF behavior will follow theory initially proposed by \citet{1976MNRAS.175..395B} in which a radiation-dominated shock forms at a radius scaling approximately linearly with the observed luminosity. We expect the imprint radius somewhere higher than the first photon scattering surface, which corresponds to a photon travelling one optical depth in the downward (parallel) direction from the top of the column. We also anticipate a negative correlation between the cyclotron centroid energy, $E_{\rm{cyc}}$, and the source luminosity, $L_{\rm X}$ \citep{BeckerEtal2012}. The shock radius in the high state should be at a higher altitude (meaning $E_{\rm{cyc}}$ should be a lower value) when compared with the low state.

\cite{Nagel1980} theorized CRSF formation is concentrated along the column where the ratio of the extraordinary photon scattering length to the absorption length is at a minimum value. He found the relationship between the resonant scattering mean free path, $\ell_{\rm scatt}$, and the absorption mean free path, $\ell_{\rm abs}$, evaluated at the centroid energy, $E_{\rm cyc}$, for fully-ionized hydrogen gas is given by
\begin{equation}
    \label{eqn:EcycRadiusNagel}
    \frac{\ell_{\rm{scatt}}}{\ell_{\rm{abs}}}=\frac{3\pi}{2} \frac{\alpha \hbar^3 c^4 n_e}{{E^3_{\rm{cyc}}}}
    \sqrt{\frac{m_e}{k_{\rm B} T_e}}\ ,
\end{equation}
where $\alpha$ is the fine structure constant, $\hbar$ is the reduced Planck constant, $k_{\rm B}$ is Boltzmann's constant, $c$ is the speed of light, and $m_e$, $n_e$, and $T_e$ denote the electron mass, number density, and temperature, respectively. The ratio in Equation (\ref{eqn:EcycRadiusNagel}) is a function of the radius, $r$, via the quantities $E_{\rm cyc}$ and $T_e$, and the radius corresponding to the minimum value of this ratio is interpreted as the cyclotron imprint radius, $r_{\rm cyc}$.

More recent results by \citet{LoudasEtAl2024} place the CRSF within the radiative shock, at a radius higher in the column than the location suggested by \cite{Nagel1980}. \citet{LoudasEtAl2024} implemented a cylindrical geometry Monte Carlo radiative transfer code and demonstrated prominent CRSFs in the emergent spectrum are generated in the shock. They also noted the possibility of evenly spaced cyclotron lines due to resonant scattering in multiple harmonics. We follow the approach of \citet{LoudasEtAl2024} here.

The {\tt WWB17} model formation of the CRSF absorption-like feature in the X-ray spectrum is described in an approximate manner by multiplying the escaping phase-averaged spectrum by a 2D Gaussian distribution, $A_{\rm cyc}(r,\epsilon)$, which depends on the radius, $r$, and the photon energy, $\epsilon$, given by
\begin{equation}
A_{\rm cyc}(r,\epsilon) = 1-\frac{d_{cr}}{2 \pi \sigma_{\rm cyc} \sigma_r}
e^{-\frac{(\epsilon - E_{\rm cyc})^2}{2 \sigma_{\rm cyc}^2}}
e^{-\frac{(r - r_{\rm cyc})^2}{2 \sigma_r^2}} \ ,
\label{eqn:cyc_abs}
\end{equation}
where $d_{cr}$ represents a combined strength parameter, and the standard deviations in the energy and radial directions are denoted by $\sigma_{\rm{cyc}}$ and $\sigma_r$, respectively. We note the cyclotron energy, $E_{\rm cyc}$, is a function of $r$ via Equation~(\ref{eqn:cycenergy}).

\subsection{Free Parameter Selection and Model Convergence}

Table~\ref{tab:free parameters} lists the values for the twelve free parameters used to model the outburst low and high states. It also includes values for three fixed quantities, three observational quantities, and several post-processing diagnostic quantities. 

We fix the NS mass and radius quantities using the canonical values $M_* = 1.4\,M_\odot$ and $R_* = 10\,$km, respectively. The third fixed quantity is established by setting the perpendicular scattering cross-section equal to the Thomson limit, $\sigma_\perp=\sigma_{_{\rm T}}$ \citep{1981A&A....93..255W}. The previously published bolometric values of $L_{\rm X}$ (at source distance $D = 55\,$kpc) are used to calculate the corresponding values of the mass accretion rate, $\dot M$, using Equation~(\ref{eq:mdot}). Hence, we classify $D, \dot M$, and $L_{\rm X}$ as the three observational quantities. 

The next step in the {\tt WWB17} modeling process, after establishing the three fixed quantities and the three observational quantities, is to select the twelve free parameters. The first six free parameters are associated with the calculation of the continuum X-ray spectrum and the dynamical structure of the AC. These comprise (1) the electron scattering cross section for photons propagating parallel to the local magnetic field direction, relative to the Thomson limit, $\sigma_{||}/\sigma_{_{\rm T}}$, (2) the angle-averaged electron scattering cross section, relative to the Thomson limit, $\bar\sigma/\sigma_{_{\rm T}}$, (3) the inner polar cap radius at the stellar surface, $\ell_1$, (4) the outer polar cap radius at the stellar surface, $\ell_2$, (5) the incident radiation Mach number at the top of the column, ${\mathscr M}_{\rm r0}$, and (6) the surface magnetic field strength, $B_*$, at the base of the column.

The final six free parameters are used to modify the spectrum during post-processing calculations, which occur after the radiative hydrodynamic equations are solved and temperatures are converged. There are five free parameters associated with dual CRSFs located at the fundamental and first harmonic cyclotron frequencies ($n=1$ and $n=2$, respectively), and one free parameter associated with interstellar absorption ($N_{\rm H}$).

Separate values for the free parameters in Table~\ref{tab:free parameters} were obtained for the low and high states, respectively, with the exception of the surface magnetic field strength, $B_*$, which is a free parameter, but was constrained to have the same value in both states, as physically required. See \citet{WestEtal2017a,WestEtal2017b} for a full discussion.

\begin{figure}
    \centering
    \includegraphics[width=\linewidth]{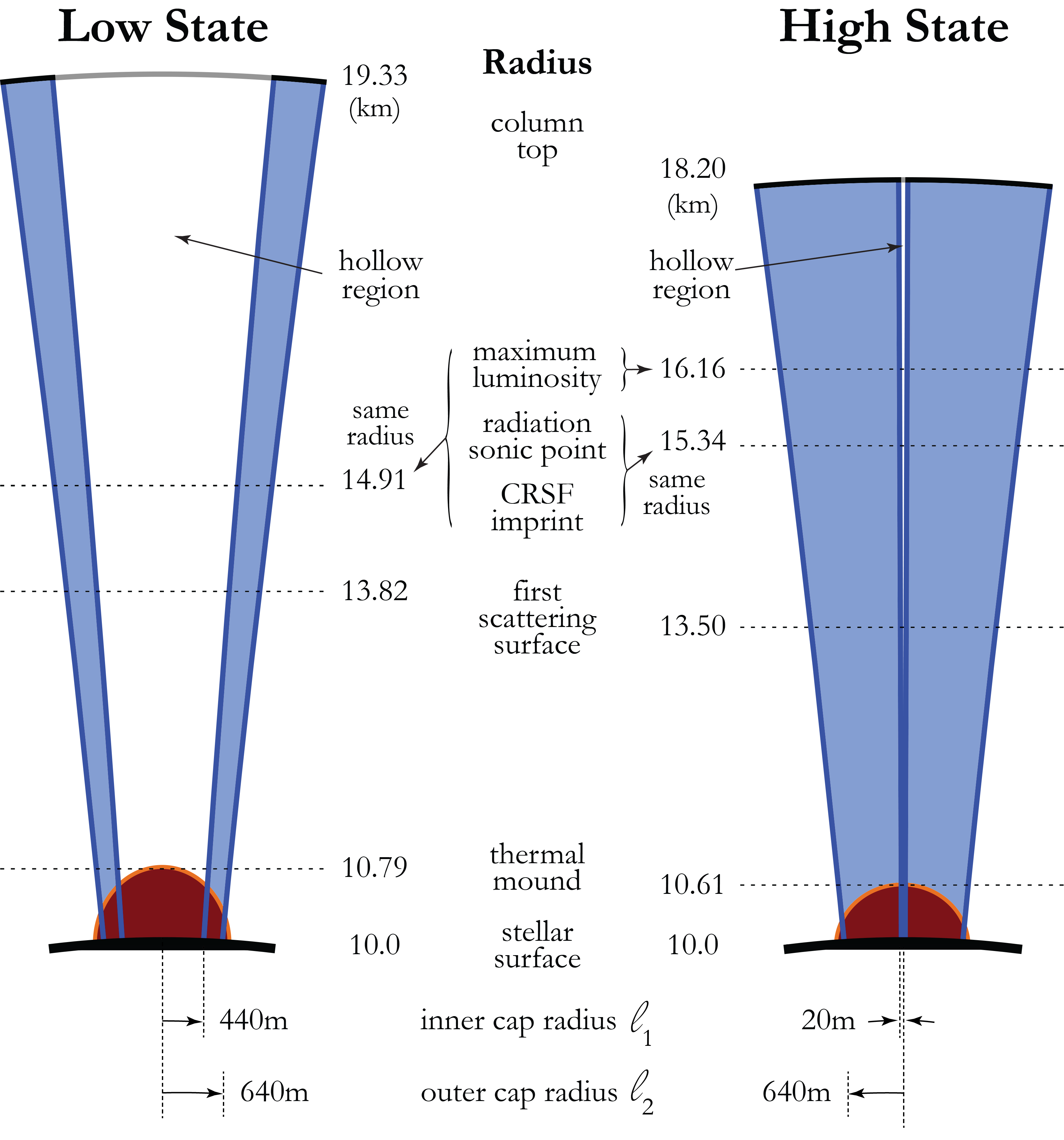}
    \caption{Cutaway view of the structure of the accretion column in the low and high states during the giant outburst of RX\,J0209.}
    \label{fig:column structure}
\end{figure}

\begin{figure}
    \centering
    \includegraphics[width=\linewidth]{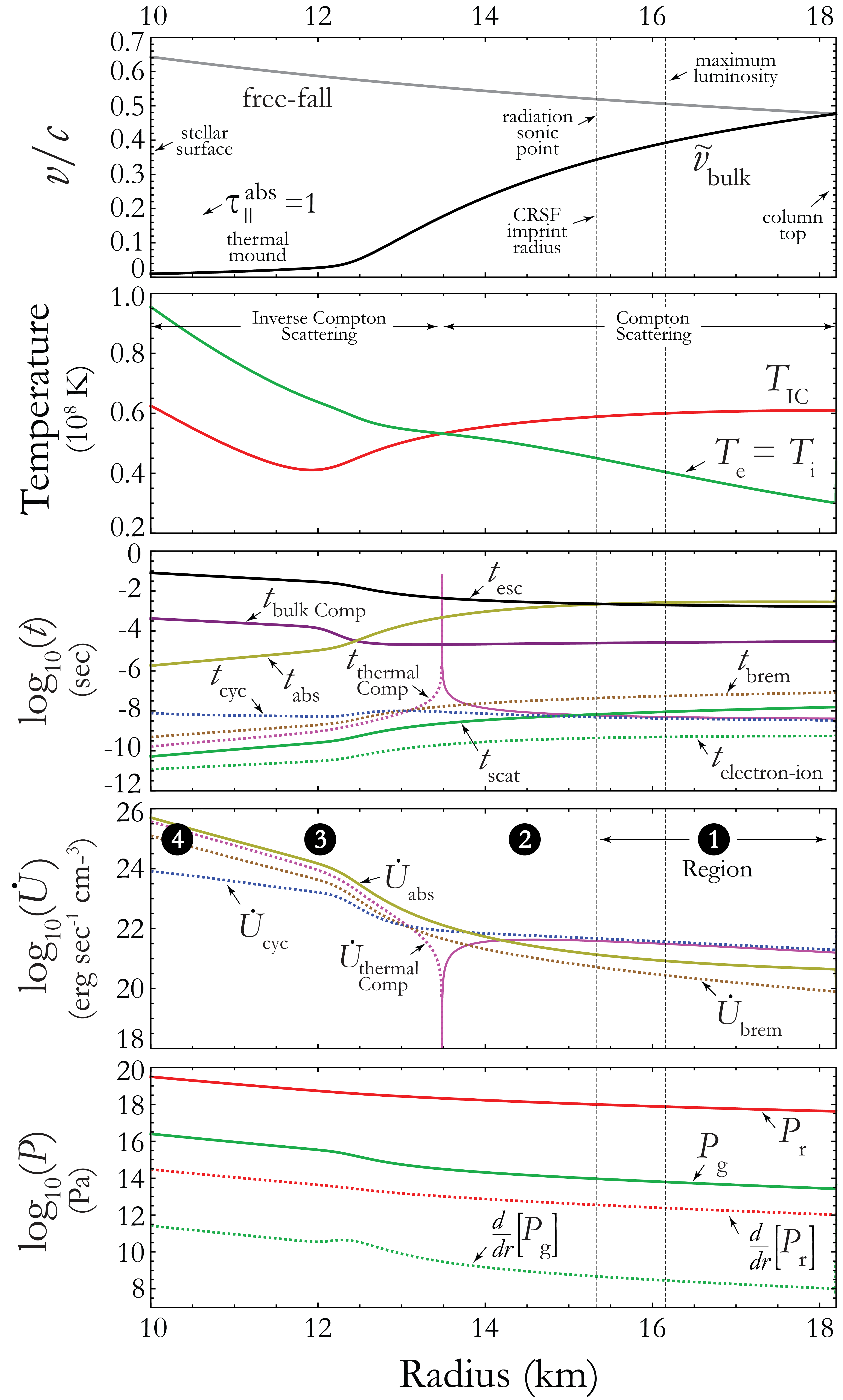}
    \caption{The model solution for the high state of the giant outburst reveals the hydrodynamic and radiative structure along the accretion column length. Timescales and energy transfer rates represent electron heating (solid curves) or electron cooling (dotted curves).}
    \label{fig:column profiles}
\end{figure}

Cutaway views of the corresponding column structures are depicted in Figure~\ref{fig:column structure}. The profiles of the primary hydrodynamic and radiative variables in the high state are plotted in Figure \ref{fig:column profiles}. Similar results were obtained in the low state; therefore, the low state profiles are not included here. The plots include the relevant timescales and energy transfer rates for the electron heating (solid curves) and cooling (dotted curves) processes.

\begin{figure*}
    \centering
    \includegraphics[width=\linewidth]{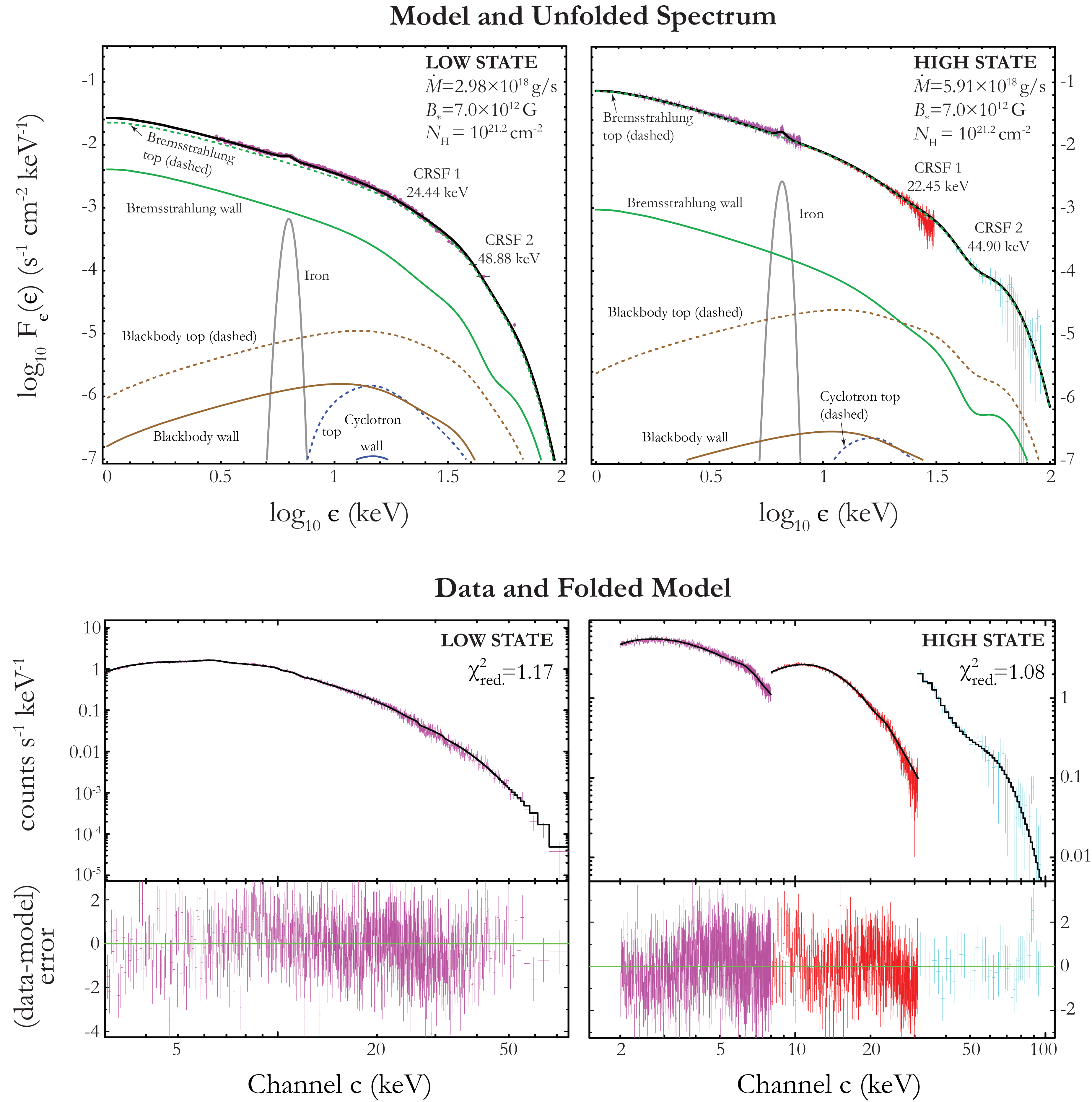}
    \caption{Theoretical spectra for the low and high states occurring in the binary X-ray pulsar RX~J0209.6-7427 during the 2019 super-Eddington outburst. {\emph{Top panel:} The various components of the phase-averaged spectra computed using the {\tt WWB17} model along with the unfolded data. The solid black lines denote the total phase-averaged spectrum, including cyclotron absorption and interstellar absorption. The solid green, brown, and blue lines in the upper panels represent the reprocessed bremsstrahlung, blackbody, and cyclotron emission components, respectively, which escape through the outer column wall. The corresponding  dashed lines indicate the components escaping through the column top. \emph{Bottom panel:} 
    The folded model with the data in the detector plane, along with the corresponding residuals and reduced $\chi^2$ values computed within XSPEC. The left panel (low state) includes the \textit{NuSTAR} data (magenta), and the right panel (high state) includes the \textit{Insight}-HXMT data for the LE (magenta), ME (red), and HE (cyan) detectors.}}
    \label{fig:phase-averaged spectra}
\end{figure*}

Finding phase-averaged spectra solutions with the {\tt WWB17} model requires connecting two software platforms in a manual fashion. The hydrodynamic equations are solved using Mathematica\footnote{https://www.wolfram.com/mathematica/}, and the photon transport equation is solved using COMSOL\footnote{https://www.comsol.com/}. An automated process, such as the one implemented in XSPEC, would require creation of a tabulated model. Each iteration of a model takes 10\,\textendash\,15 minutes; hence, this approach would require years of computation for a reasonably dense grid when considering all twelve free parameters. Therefore, automated fitting in XSPEC is not practical at this time. The model parameters were simply varied manually. The resulting theoretical spectra were then folded through the detector response matrix using XSPEC to make a quantitative comparison with the observed data in the detector plane. Using this method, we were able to quantitatively compare the {\tt WWB17} model with the observed data, yielding values for the reduced $\chi^2$, as well as plots of the residuals. We find that $\chi^2_{\rm red.} = 1.17$ in the low state and $\chi^2_{\rm red.} = 1.08$ in the high state. These results are included in Figure~\ref{fig:phase-averaged spectra}.

\begin{deluxetable}{lll}
\tablewidth{0pt}
\tablecaption{Model values for the low and high states.}
\label{tab:free parameters}
\tablehead{\colhead{Free Parameters} & \colhead{Low State} & \colhead{High State}}
\startdata
Parallel cross-section $\sigma_\parallel/\sigma_{_{\rm T}}$ & $4.44\times10^{-4}$ & $3.88 \times 10^{-4}$ \\
Angle-averaged cross-section $\bar{\sigma}/\sigma_{_{\rm T}}$ & $16.0\times10^{-4}$ & $9.50 \times 10^{-4}$ \\
Outer polar cap radius $\ell_2$ (m) & 640 & 640 \\
Inner polar cap radius $\ell_1$ (m) & 440 & 20 \\
Incident radiation Mach number ${\mathscr M}_{r0}$ & 1.66 & 1.40 \\
Surface magnetic field $B_{*}$ ($10^{12}\,$G) & 7.0 & 7.0 \\
CRSF radial $\sigma_{r}$ (km) & 4.1 & 4.1 \\
CRSF 1 energy $\sigma_{\rm cyc}$ (keV) & 11.0 & 8.25 \\
CRSF 1 strength $d_{cr}$ (km\,keV)  & 95.0  & 52.0 \\
CRSF 2 energy $\sigma_{\rm cyc}$ (keV) & 16.0 & 16.5 \\
CRSF 2 strength $d_{cr} ({\rm km\,keV}) $ & 195.0  & 202.0 \\
%
$N_{\rm H}$ ($10^{21}$cm$^{-2}$) & 1.58 & $1.58$ \\
\hline
\colhead{Fixed Quantities} & \colhead{Low State} & \colhead{High State} \\
\hline
Perpendicular cross-section $\sigma_\perp/\sigma_{_{\rm T}}$ & $1.0$ & $1.0$ \\
$M_*$ ($M_\odot$) & 1.4 & 1.4 \\
$R_*$ (km) & 10 & 10 \\
\hline
\colhead{Observational Quantities\textsuperscript{\textdagger}} & \colhead{Low State} & \colhead{High State} \\
\hline
$L_{\rm X}$ (erg s$^{-1}$) & $5.54\times10^{38}$ & $1.10\times10^{39}$ \\
$\dot M$ (g s$^{-1}$) & $2.98 \times 10^{18}$ & $5.91 \times 10^{18}$ \\
Distance $D$ (kpc) & 55 & 55 \\
\hline
\colhead{Diagnostic Quantities\textsuperscript{\textdaggerdbl}} & \colhead{Low State} & \colhead{High State} \\
\hline
$L_{\rm X}^{\rm tot}$ (erg s$^{-1}$) & $5.06\times10^{38}$ & $1.03\times10^{39}$ \\
$L_{\rm X}^{\rm brem}$ (erg s$^{-1}$) & $5.02\times10^{38}$ & $1.02\times10^{39}$ \\
$L_{\rm X}^{\rm cyc}$ (erg s$^{-1}$) & $2.52\times10^{35}$ & $4.32\times10^{34}$ \\
$L_{\rm X}^{\rm bb}$ (erg s$^{-1}$) & $3.46\times10^{36}$ & $7.59\times10^{36}$ \\
Sonic point radius (km) & 14.91 & 15.34 \\
CRSF imprinting radius $r_{\rm cyc}$ (km) & 14.91 & 15.34 \\
CRSF 1 energy $E_{\rm cyc}$ (keV) & 24.44 & 22.45 \\
CRSF 2 energy $E_{\rm cyc}$ (keV) & 48.88 & 44.90 \\
Radius of max wall emission (km) & 14.91 & 16.16 \\
First (parallel) scattering surface (km) & 13.77 & 13.38 \\
%
$L_{\rm top}$ (erg s$^{-1}$) & $4.48\times10^{38}$ & $1.02\times10^{39}$ \\
$L_{\rm wall}$ (erg s$^{-1}$) & $5.78\times10^{37}$ & $1.13\times10^{37}$ \\
Thermal mound $T_{\rm{th}}$ (K) & $2.90\times10^{8}$ & $3.00\times10^{8}$ \\
Surface impact velocity $\upsilon_*/c$ & $8.07\times10^{-3}$ & $9.63\times10^{-3}$ \\
Incident ion Mach number ${\mathscr M}_{i0}$ & 203.0 & 183.6 \\
\enddata
\tablecomments{
\textsuperscript{\textdagger}The value of $\dot{M}$ is computed from $L_{\rm X}$ using Equation (\ref{eq:mdot}), see the text for the discussion of the bolometric luminosity $L_{\rm X}$.
\textsuperscript{\textdaggerdbl}Diagnostic luminosities are computed in the standard energy range 3-79 keV energy for both states for purposes of comparison.
}
\end{deluxetable}

\section{Discussion} \label{sec:discussion}

The good agreement between the {\tt WWB17} theoretical model and the observed data obtained in the low and high states of the 2019 giant outburst allows us to make quantitative comparisons of the model parameters and physical profiles in the two states. Overall, the radiation hydrodynamics, energy exchange rates, and timescales are very similar in the two states, but there are some interesting differences.

Figure~\ref{fig:column profiles} depicts the radiative hydrodynamic profiles along the full length of the AC for the high state converged model solution. The five panels in descending order depict the bulk velocity, temperatures, timescales, energy coupling rates, and pressure and pressure gradients. The timescales and energy coupling rates associated with gas heating processes are indicated by solid curves, whereas gas cooling processes are indicated by dotted curves. We segment the 8.2\,km length of the column into four regions, which are labeled in the $4^{\rm th}$ panel of Figure \ref{fig:column profiles} (energy coupling rates). The regions separate hydrodynamic milestones as the bulk fluid enters the top of the column and eventually stagnates at the stellar surface and merges with the NS crust.

Region 1 begins where the accreting matter enters the column top with a velocity of $0.48\,c$. The high gas density ensures the electrons and ions equilibrate to the same temperature over the entire length of the column. Cyclotron cooling dominates all energy coupling rates in Region 1 on a timescale shorter than any heating rate, and therefore the gas temperature does not rapidly equilibrate (higher in the column) to the inverse-Compton temperature of the radiation field, $T_{\rm IC}$. The radius of maximum wall emission is at 16.2\,km as photons continue transferring energy to the electrons via Compton scattering while approaching the radiation sonic point.

The plasma enters Region 2 after passing through the radiation sonic point while continuing to decelerate. The thermal Comptonization gas heating rate is a maximum at the top of Region 2, and starts dropping rapidly as photons and electrons approach the same temperature at the bottom of Region 2. The free-free absorption rate quickly rises and becomes the dominant gas heating mechanism. Electrons and photons finally reach thermal equilibrium at the inverse-Compton temperature, $T_{\rm IC}$, which marks the end of Region 2. The Compton scattering process transitions into inverse-Compton scattering downstream from this radius, as the accreting gas enters Region 3, where much of the remaining electron energy is transferred to the photons.

Free-free absorption continues to dominate in Region 3 as the electron temperature increases rapidly. Radiation pressure continues to decelerate the material as it transitions into the sinking regime discussed by \citet{1976MNRAS.175..395B}. Region 4 begins at the top of the thermal mound, where the parallel absorption optical depth, $\tau^{\rm{abs}}_{\parallel}$, equals unity. The electrons and protons decelerate to rest inside the thermal mound and merge with the star's crust.

We follow \cite{LoudasEtAl2024} and set the CRSF imprint radius equal to the radiation sonic point radius, and we allow for the possibility of dual CRSFs appearing at both the fundamental cyclotron frequency ($n=1$), as well as the first harmonic frequency ($n=2$). Multiple harmonics are reported in the literature for other sources. For example, \citet{2020A&A...634A..99B} describe a model with a set of five harmonics for the 1999 giant outburst of 4U 0115+63. In our simulations, the centroid energy for the fundamental frequency was calculated to be 24.44\,keV and 22.45\,keV for the low and high states, respectively. The centroid energy for the first harmonic was set to twice the fundamental value, corresponding to energies 48.88\,keV and 44.90\,keV, respectively.

One of the most important free parameters is the surface magnetic field strength at the polar cap, for which we obtained $B_* = 7.0 \times10^{12}$ G, a value that can be tested against observational constraints. The best observational estimates of the \rxjs surface $B$-field strength rely on two methods. These include the application of torque models that can explain the spin evolution of the NS during the 2019 outburst, and the identification of critical transitions in either the Hardness Ratios (HR) or the pulse profile shape that are believed to be related to changes in column morphology.

Torque modeling of the \rxjs data, assuming a distance of $D = 55\,$kpc, reveals a magnetic field strength of 2\,\textendash\,6$\times10^{12}\,$G at the NS poles, with uncertainties mainly due to the adopted bolometric correction and the details of the specific torque model \citep[e.g.,][and discussions therein]{2023MNRAS.520..281K,2022ApJ...938..149H}. On the other hand, critical transitions in the pulse profile shape \citep{2022MNRAS.517.3354L} and HRs \citep[see][and also our Figure~\ref{fig:LC}]{2022ApJ...938..149H} suggest a higher $B$-field strength of $\sim2\times10^{13}$~G. The differences between the two estimates could be explained by the presence of a multipolar $B$-field \citep{2022MNRAS.517.3354L}.

An additional important result of our analysis is an improved understanding of how the structure of the accretion column evolves with increasing luminosity during the outburst. The outer column radii in both states is found to be 640\,m. The unchanged outer polar cap radius is expected because the inner edge of the radiation-dominated accretion disk should not vary significantly, if at all, with the increased mass flow rate \citep{2017MNRAS.470.2799C}. The pressure balance between the disk and magnetic field in this inner region remains essentially constant. However, the column is significantly more hollow in the low state (440\,m inner polar cap radius), as compared with the high state (20\,m inner polar cap radius), which is nearly completely filled. See Figure~\ref{fig:column structure}.

More puzzling is the transition from a mostly hollow column in the low state to a nearly filled column in the high state. The inner polar cap radius obtained in the low state is $\ell_1 = 440\,$m, whereas in the high state we obtained  $\ell_1 = 20\,$m. This question can't be fully resolved without incorporating a detailed model for the structure of the extended pulsar magnetosphere and the surrounding accretion disk, including the binary orbit and the properties of the companion star, which is beyond the scope of this paper.

We believe that the changes in the inner polar cap radius, $\ell_1$, between the low and high states are due to the super-Eddington accretion rates involved here. The high luminosities during the outburst will undoubtedly cause some heating of the accretion disk, which may cause it to puff up \citep[e.g.,][]{2017MNRAS.470.2799C,Chashkina2019,Mushtukov2019,MushtukovEtal2024}. This change in the accretion disk geometry will cause a large variation in the radius where the matter is first picked up and entrained onto the magnetic field. In particular, the outer-most pick-up radius is likely to move outwards as the disk puffs up in the high state. Keeping in mind that the outer-most pick-up radius corresponds to the inner wall of the accretion column where it connects with the stellar surface, we conclude an increase in the accretion rate will tend to produce a more filled column in the high state, as we obtained here.

\section{Conclusion}

In this study we modeled the spectral formation occurring in the binary X-ray pulsar RX J0209.6-7427 during the 2019 super-Eddington Type~II outburst. We focused on analysis of the X-ray spectra detected during low and high states, between which the accretion rate changed by a factor of $\sim 2$. We employed the previously published self-consistent {\tt WWB17} model for the radiative and dynamical processes occurring in the AC of the highly magnetized NS, and we find that the simulated spectra closely match the observed spectra during the low and high states, observed with {\it NuSTAR} and {\it Insight}-HXMT, respectively.

We review our primary findings here and focus on a comparison of the free parameters and the radiative and hydrodynamic profiles inside the AC. This reveals a number of interesting changes as the outburst transitions from the low state to the high state:

\begin{enumerate}[1)]
    \item Bremsstrahlung photons provide the primary source of seed photons contributing to the phase-averaged spectrum. The cyclotron and blackbody seed photons make a negligible contribution. The luminosities of the various respective reprocessed seed photon components are indicated in Table~\ref{tab:free parameters}. It is clear that bremsstrahlung provides over 99\% of the power in the observed spectra for both the low and high states.
    
    \item The luminosity radiated from the column top is significantly higher than the wall luminosity for both states, and the relative contribution of the top component increases with luminosity (see Figure \ref{fig:phase-averaged spectra} and Table~\ref{tab:free parameters}). This has immediate implications for the shape of the resulting pulse profile. For example, we predict the pulsar profiles of the ULXPs should be roughly sinusoidal, as observed in many cases \citep[e.g., NGC 300 ULX1;][]{Carpano2018}.

    
    \item The angle-averaged and parallel electron scattering cross sections, $\bar\sigma$ and $\sigma_{||}$, respectively, are larger in the low state. The enhanced cross sections in the low state are due to the stronger cyclotron absorption parameter, $d_{cr}$, at the fundamental frequency ($n=1$) which amplifies the effect of the cyclotron resonance, hence increasing the scattering cross sections (see Table~\ref{tab:free parameters}). The reduced cross sections in the high state cause a larger fraction of the photons to be beamed along the magnetic field lines, leading to a larger column top luminosity. Our findings are consistent with the results of \citet{Ventura1979} and \citet{MeszarosAndVentura1979}.
    
    \item The CRSF energy, $E_{\rm cyc}$, is 2\,keV lower and the radiation sonic point is 0.43\,km higher in the high state when compared with the column structure in the low state. This expected behavior is in agreement with the ideas first proposed by \cite{1976MNRAS.175..395B}. The theoretical model developed by \cite{BeckerEtal2012} predicts a negative correlation between centroid energy and emission altitude according to
    \begin{equation}
        \frac{E_{\rm cyc}}{E_*}=\left(\frac{R_*+h}{R_*}\right)^{-3} ,
        \label{eqn:Becker_eqn58}
    \end{equation}
    where $E_*$ is the surface value for the cyclotron energy according to Equation (\ref{eqn:cycenergy}). The $h$ parameter corresponds to the emission altitude at which photon diffusion through the column walls is a maximum. Setting the value of $h$ equal to the CRSF imprint altitude (which is also the sonic point altitude) establishes that Equation (\ref{eqn:Becker_eqn58}) is essentially a combination of Equations (\ref{eqn:cycenergy}) and (\ref{eqn:BfieldStrength}).
    
    \item The super-Eddington luminosities in Type~II outbursts do not disrupt the accretion column structure \citep[][]{MushtukovEtal2024} because the scattering cross section for photons propagating parallel to the magnetic field is sub-Thomson by several orders of magnitude \citep{1981A&A....93..255W}. There can be an indirect effect on the AC structure, however, if the super-Eddington flux heats the surrounding accretion disk, thereby causing changes in the accretion stream feeding into the column. \citep[e.g.,][]{2017MNRAS.470.2799C,Chashkina2019}.
\end{enumerate}


Our results have important implications for the application of AC radiative models to the analysis of major outbursts from BeXRBs and ULXPs. In particular, ULXPs share spectral similarities with BeXRBs \citep[e.g.,][]{2017A&A...608A..47K,2018ApJ...856..128W}, which have so far mainly been explored using empirical and phenomenological models. Next-generation X-ray telescopes, such as the High Energy X-ray Probe \cite[HEX-P;][]{2023arXiv231204678M}, will have higher effective areas extending over larger energy ranges than current detectors, which will facilitate much more detailed studies of super-Eddington bursts \citep[see simulations by][]{2023FrASS..1092500L,2023FrASS..1089432B}.

\begin{acknowledgments}
We are grateful to the anonymous referee, who made several important comments and suggestions that led to substantial improvements in the manuscript. GV acknowledges support by H.F.R.I. through the project ASTRAPE (Project ID 7802). The authors would also like to thank Dr. Kenneth Wolfram for his insights and constructive conversations.
\end{acknowledgments}

\bibliography{general}{}
\bibliographystyle{aasjournal}

\end{document}